# DUTCH COLONIAL TIME: TIME SIGNALS IN PARAMARIBO AND THE DUTCH CARIBBEAN

**Richard de Grijs**
*School of Mathematical and Physical Sciences, Macquarie University,
Balaclava Road, Sydney, NSW 2109, Australia*
Email: richard.de-grijs@mq.edu.au

**Abstract:** In the nineteenth century, the Dutch established time signals in their Atlantic colonies to synchronise maritime navigation with European standards. In Paramaribo (Suriname), a sophisticated sequence of apparatus—including time balls, noon guns, discs and flags—operated from 1851 until World War I. Naval officers aboard guard ships used sextants equipped with artificial horizons to determine local noon, thus integrating the colony into the global Greenwich-based cartographic system. This infrastructure was not merely technical; it became a civic ritual, with the daily noon gun structuring urban life and becoming a point of political negotiation between naval commanders and the colonial governor. In contrast, the Dutch Caribbean islands employed simpler, pragmatic systems. Curaçao used a daily time flag, a cost-effective solution suited to its climate and harbour scale, while smaller islands like Aruba and St. Eustatius relied on occasional noon guns. This diversity reflected a decentralised colonial administration that adapted technologies to local conditions and budgets. The history of these time signals reveals a process of hybrid adaptation, not simply replication of European models. They were shaped by environmental challenges, fiscal constraints and local politics, functioning simultaneously as navigational aids and civic landmarks. Their eventual decline, owing to budgetary pressures and new technologies like wireless telegraphy, underscores the fragile and negotiated nature of colonial scientific infrastructures.

## 1 INTRODUCTION

The dissemination of precise time was one of the defining markers of nineteenth-century maritime modernity. Ships' chronometers, marine surveys, harbour infrastructure and civic rhythms all depended on an increasingly exact regime of temporal coordination. By the 1850s, time balls—simple devices in which a large sphere was hoisted and dropped at a specified instant—had become familiar fixtures in major ports across Europe and its colonies. Their daily descent at a specified time served simultaneously as a public punctuation of civic life and as a vital calibration point for chronometers on board merchantmen and naval vessels. In Britain, France and the United States, the installation of time balls, accompanied in many cases by the firing of noon guns, testified to the growing global reach of a 'temporal discipline' (Thompson, 1967; see also, e.g., Glennie and Thrift, 1996; Landes, 1983) that shaped seafaring, commerce and everyday life alike.

For the Netherlands, a mid-sized maritime power by the nineteenth century, the challenge was to extend this infrastructure to her colonies. Dutch ports at home, such as Den Helder, Hellevoetsluis and Vlissingen, were quickly equipped with time balls, linked to metropolitan observatories and telegraphic time distribution (Mohrmann, 2003). Yet the colonial empire presented distinctive opportunities and difficulties. Time signals were indispensable for vessels navigating the shallow coasts of the Dutch South American colony of Dutch Guiana or Surinam (now Suriname), for fleets departing from the islands of Curaçao or St. Eustatius in the Caribbean or for ships bound across the Atlantic or around the Cape of Good Hope. At the same time, the adaptation of European practices to colonial environments was uneven and sometimes controversial. Local hydrographic officers had to deal with tropical climates, limited budgets and the political authority of governors, while also integrating their work into global cartographic systems based increasingly on the Greenwich meridian as zero-longitude marker.

The historiography of time signals has until recently concentrated on European centres, especially the Royal Observatory at Greenwich, whose time ball (erected in 1833) became the global archetype (e.g., Bartky and Dick, 1981; Gillin, 2020; Howse, 1980; Kinns, 2022). Historians of navigation have emphasised how precision chronometry combined with visual time signals was harnessed to make oceanic travel safer and more reliable (Howse, 1980). Historians of science and empire, meanwhile, have extensively explored the colonial role of botanical gardens, meteorological observatories and medical stations, but rarely the importance of the local time ball (e.g., Raj, 2007). Yet, colonial time signals reveal much about the circulation of technology, the transfer of expertise and the embedding of European norms within colonial societies.

Here I focus on two little-studied cases in the context of time signals: the Surinamese capital, Paramaribo, and the Dutch Caribbean islands. In Paramaribo, a sequence of apparatuses—time balls, guns, discs and flags—offered sailors a means to verify their chronometer settings for more than half a century, from 1851 until World War I. In the Caribbean, meanwhile, signals took simpler forms, often reduced to flags or gunfire, reflecting the smaller scale of local harbours and the influence of regional practice. Together, these signals illustrate how Dutch colonial authorities exported European scientific infrastructures, while also adapting them to local conditions and embedding them in civic life.

First, I will examine the material apparatus of colonial time signals. Unlike the Netherlands itself, where debates over balls versus discs (often referred to as 'flaps'; see my companion article) engaged astronomers such as Frederik Kaiser (1808–1872), colonial systems were shaped by pragmatic concerns: the availability of guard ships, the visibility of forts or the willingness of governors to sanction expenditure. In Paramaribo, the apparatus migrated from ship to shore, from ball to disc to flag, in a continuous process of adaptation. Curaçao, by contrast, only maintained a flag, while smaller Dutch Caribbean islands relied on infrequent noon guns to punctuate local noon.

Second, I will explore the social and political dimensions of these signals. In Suriname, disputes between naval commanders and Governor Cornelis Ascanius van Sypesteyn (Sijpesteyn; 1823–1892) in the 1860s reveal how questions of scientific accuracy intersected with colonial authority and civic routine. In Curaçao, the adoption of a flag signal underscored the communication priorities of a trading port whose rhythms were tied to intra-Caribbean shipping. These cases remind us that technologies of time were never purely technical; they were also political symbols and turned into civic rituals.

By emphasising developments in Paramaribo and the Caribbean, I aim to broaden the geography of Dutch time-signal history. I show how the Dutch empire's Atlantic wing participated in the same temporal transformations as Amsterdam, Rotterdam and the Dutch seaports, albeit on different terms. The story of Dutch colonial timekeeping complicates simple narratives of diffusion from European observatories. It reveals instead a patchwork of hybrid infrastructures, entangled with local conditions and imperial hierarchies, projecting European authority even as they were reshaped in tropical ports.

## 2 PARAMARIBO: TIME BALLS, GUNS, DISCS AND FLAGS

Paramaribo was the principal Dutch port on the Guianas coast, serving as a linchpin between Europe, the Caribbean and the plantation hinterland. From 1851 onwards, the town operated a time-signal system that, although modest in scale compared to the other main Dutch colonial outpost in Batavia (Jakarta, Indonesia; e.g., Kinns, 2021; Orchiston et al., 2021), embodied the core features of Dutch maritime timekeeping. It evolved through a sequence of devices—guard-ship time balls, shore-based guns, discs, flags and cylinders—which together projected European expertise into colonial space while also acquiring distinctive local meanings.

### 2.1 Beginnings: the *Heldin* and the first time ball (1851–1853)

The origins of Paramaribo's time signals are found in a naval initiative of 1851. On 15 November that year, the Dutch Royal Navy's corvette *Heldin* ('Heroine'), serving as guard ship in the Paramaribo roadstead, was equipped with a time ball. Operated from the foremast, the device was hoisted in advance and dropped at noon, local mean time. Commencement of time-signal operations was formally announced in the Surinamese national newspaper (Wesenhagen, 1851; own translation):

> Government Secretariat.
> Paramaribo, 12 November 1851.
>
> Notice to Mariners.
>
> Following a communication received from the Captain at Sea, commander of His Majesty's Naval Forces in the West Indies, it is hereby made known—by order of His Excellency the Governor of this colony—to seafarers that, beginning on the 15th of this month, a time ball will be dropped from the rigging of H.M. corvette *Heldin*, exactly at the moment of local mean noon. Accordingly, it must be considered to be precisely 12 o'clock (mean time) at the instant the ball begins to fall.
>     As a warning, the time ball on the said ship will be hoisted halfway up the mast five minutes before mean noon, and all the way to the top two minutes before mean noon.[1]
>     For information, it is noted that the difference in longitude and in time of the naval dock (which is nearly identical with the location of this time ball) compared with the meridian of Greenwich is on average:
>
> in longitude = 55° 12′ 34″
> in time = 3h 40m 51s.

In the absence of the presiding Government Secretary,
The Chief Clerk,
J[acob]. E[evert]. Wesenhagen [1813–1875]

This was the first attempt to replicate European practice in the Dutch Atlantic. Unlike in the Netherlands, there was no observatory ashore to provide precise determinations. Instead, naval officers relied on sextant observations of the Sun's meridian passage, using artificial horizons to compensate for uneven ground and humid conditions. The difficulties were considerable. Suriname's climate produced sudden tropical showers and frequent cloud cover at critical moments, complicating the determination of the precise time of noon.

Hydrographers such as J. Vos,[2] Arnoldus H. Bisschop Grevelink (1811–1882) and Jacob Swart (1807–1879) combined astronomical observations with detailed surveys of the Suriname River. They also ensured that the ball dropped at the correct instant. Municipal records sometimes listed the responsible supervisors, suggesting that their role carried public recognition. That recognition and responsibility were underscored by the fact that time-ball supervisors usually held the rank of lieutenant at sea, first class (senior lieutenant rank), in the Royal Dutch Navy (Mohrmann, 2003); at early times, qualified lieutenants at sea, second class (mid-rank), were sometimes also tasked with these duties (e.g., Swart, 1853: 166): "Since the establishment of [the Paramaribo] time ball, Second Lieutenant W[illem]. F[rederik]. van Erp Taalman Kip, serving on the corvette *Heldin*, has carried out his duties particularly well", we learn from the vessel's captain, Willem Stort (1798–1867), in December 1852.

Their calculations fixed the coordinates of the guard ship at 5° 44′ 30″ N, 55° 12′ 54″ W (van Cleef, 1858) and recorded Paramaribo's mean local time as 3h 40m 35s 'slow' relative to Greenwich (Milne, 1899; van Cleef, 1858). These measurements integrated Suriname into a global cartographic order increasingly defined by Greenwich. Earlier Dutch charts had used Tenerife as their prime meridian (e.g., de Grijs, 2017: Chapter 5). By mid-century, however, the Netherlands aligned with Britain, acknowledging both scientific pragmatism and geopolitical reality. Paramaribo's ball thus synchronised not only local shipping but also Dutch imperial authority with a British-defined standard.

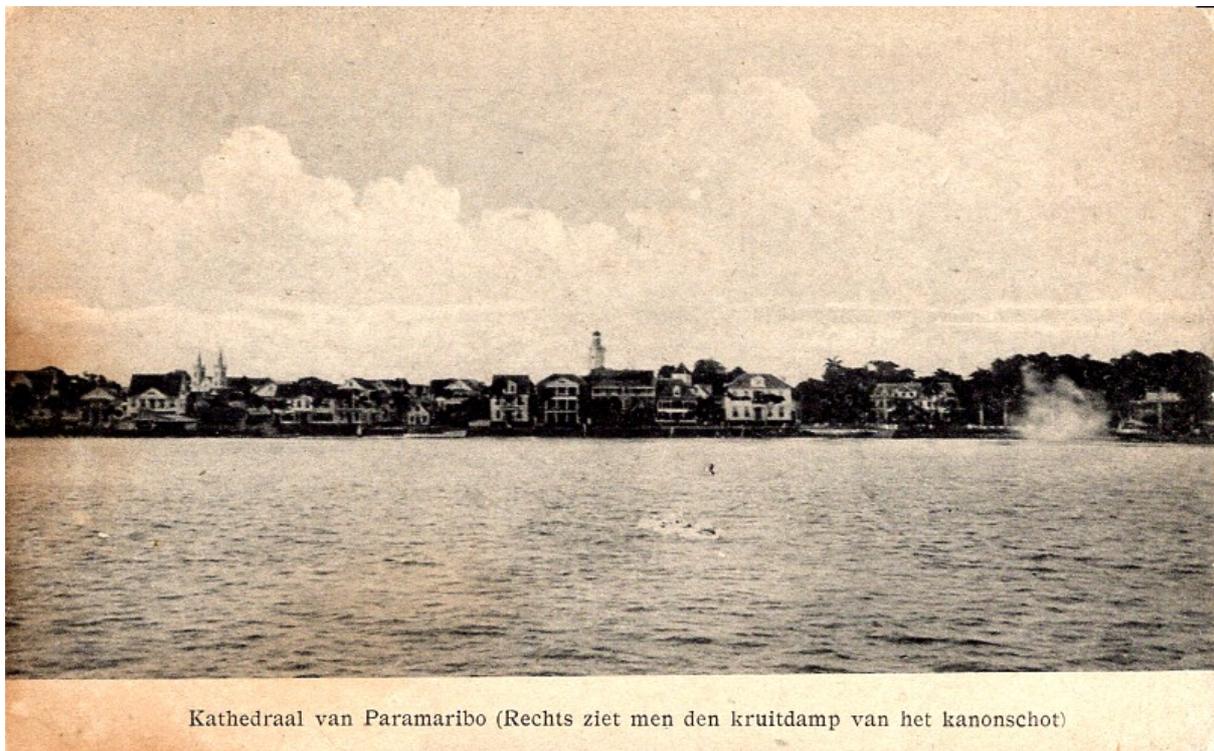

Kathedraal van Paramaribo (Rechts ziet men den kruitdamp van het kanonschot)

**Figure 1.** Vintage postcard of Paramaribo's seascape (Klaus Hülse collection). The caption reads, "Paramaribo cathedral (On the right, the gunpowder smoke from the cannon is visible)."

The time ball was tied to the daily *middagschot* (noon gun; see Figure 1). Naval regulations prescribed gunfire in the morning, at noon and in the evening; in Paramaribo, the noon shot was synchronised with the fall of the ball. This dual signal—visual and auditory—ensured that mariners could verify their chronometers even in poor visibility. Responsibility for the time signal was onerous. When the *Heldin* was absent between January and May 1852, the *Sperwer* ('Sparrowhawk') and *Koerrier* (*sic*; 'Courier') assumed the task (Swart, 1853: 166). Reports noted the difficulty of obtaining uninterrupted series of observations, but the presence of the signal was considered indispensable. Contemporary records imply that the Paramaribo time signal was sufficiently accurate to allow visiting mariners to adequately calibrate their timepieces (Swart, 1853: 166–167; own translation):

> Since the 15th of November of last year [1851], the dropping of the time ball, rigged on H.M. corvette *Heldin*, has been carried out at the mean noon of Paramaribo.
> It has been shown satisfactorily that the chronometer of HOWHU,[3] Navy No. 113, Factory No. 113, has maintained a very regular and uniformly accelerating rate, both on the outward voyage and thereafter. For that reason, this timepiece has specifically served for the falling times, and continues to give satisfaction—especially when one considers that the weather conditions often prevent, for days on end, observations for the hour angle. In normal circumstances, these observations take place every three to four days with the aid of the artificial horizon, three series of 5 or 7 altitudes usually being taken, from which the mean is derived.
> From the various verbal reports of officers of H.M. ships, as well as—among others— of the merchant captains Spekman, commanding the *Lodewijk Antonie*, Visser of the *Elisa*, and Mr. H. Lyon, owner of the *Anna*, who, touching at the colony on their homeward voyages, obtained good results, I believe I may state with confidence that this time ball meets the requirements, and consequently provides great benefit to navigation.
>
> On board H.M. corvette *Heldin*,
> at the roadstead of Paramaribo, 31 December 1852.
>
> The Captain at Sea, Commander of
> H.M. Naval Forces in the West Indies,
> W. Stort

## 2.2 Astronomical foundations of colonial time signals

At the heart of every colonial time signal lay a simple astronomical procedure: the determination of local apparent noon. This was the instant when the Sun reached its highest altitude in the sky, crossing the local meridian. By definition, this moment marked twelve o'clock local mean time. But achieving the necessary precision in a tropical colony was far from straightforward.

Naval officers relied on sextant observations, measuring the altitude of the Sun as it approached its meridian passage. Because the true horizon in Paramaribo was often obscured by vegetation or cloud banks, they employed artificial horizons—bowls of mercury or glass reflecting surfaces—to obtain accurate angles. A single observation was rarely sufficient; several had to be averaged to reduce error. Refraction tables corrected for the bending of light through the atmosphere, while the 'equation of time' adjusted for the eccentricity of Earth's orbit, which caused apparent solar noon to deviate from mean solar noon by up to a quarter of an hour, depending on the time of year.

The astronomical knowledge embedded in these procedures was both practical and scholarly. Officers drew on the *Nautical Almanac* (published in Britain since 1767) and its Dutch counterparts,[4] which tabulated the Sun's declination and the equation of time for every day of the year. By comparing their local observations to these tables, they could identify the precise moment of noon. Only then could the ball be dropped or the gun fired.

Chronometer comparisons added an additional layer of astronomical reasoning. Because longitude could be expressed as a time difference relative to a reference meridian, mariners compared their shipboard chronometers against the local signal to detect timepiece drift. A chronometer that read 12:03 at the instant of the noon gun was three minutes fast; at the equator, this implied an error of about 45 nautical miles in longitude. The seemingly minor adjustment of a few seconds could thus be the difference between safe passage and shipwreck.

In adopting Greenwich as their reference, Dutch hydrographers also embraced a broader astronomical consensus. The Royal Observatory's prestige and the international adoption of Greenwich Mean Time by hydrographic offices made it the *de facto* global standard. When officers in Paramaribo determined that local noon was 3h 40m 35s behind Greenwich, they were situating Suriname within a framework of celestial reference. The Paramaribo ball, in other words, was not only a mechanical device but also an astronomical performance mediated through naval science and colonial authority.

## 2.3 Institutionalisation and the noon gun

In the mid-1850s, the system was institutionalised. Fort Zeelandia, the seventeenth-century stronghold on Paramaribo's waterfront (see Figure 2), became the shore-based anchor. Here a cannon was fired daily at noon to accompany the drop of the ball from the guard ship. The noon gun quickly acquired civic significance (Oudschans Dentz, 1927/8). Newspapers mentioned its irregularities; residents set their watches by it; school schedules and office hours aligned with its report. The tripartite rhythm of morning, noon and evening guns structured the day far beyond maritime needs. In effect, Paramaribo's civic temporality was colonised by naval routine.

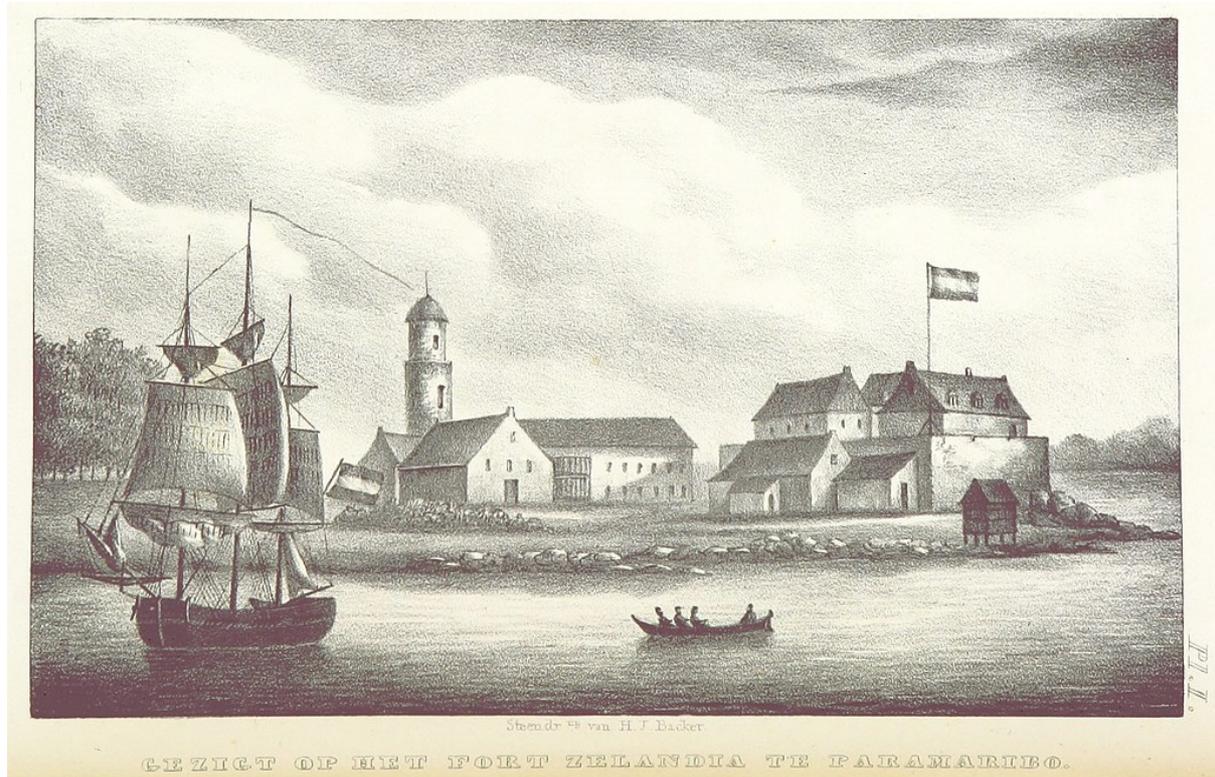

**Figure 2.** Fort Zeelandia. (G. Coster van Lennip, 1842. *Aanteekeningen, gehouden gedurende mijn verblijf in de West-Indiën, in de jaren 1837–1840*: 44. British Library HMNTS 10470.d.3.)

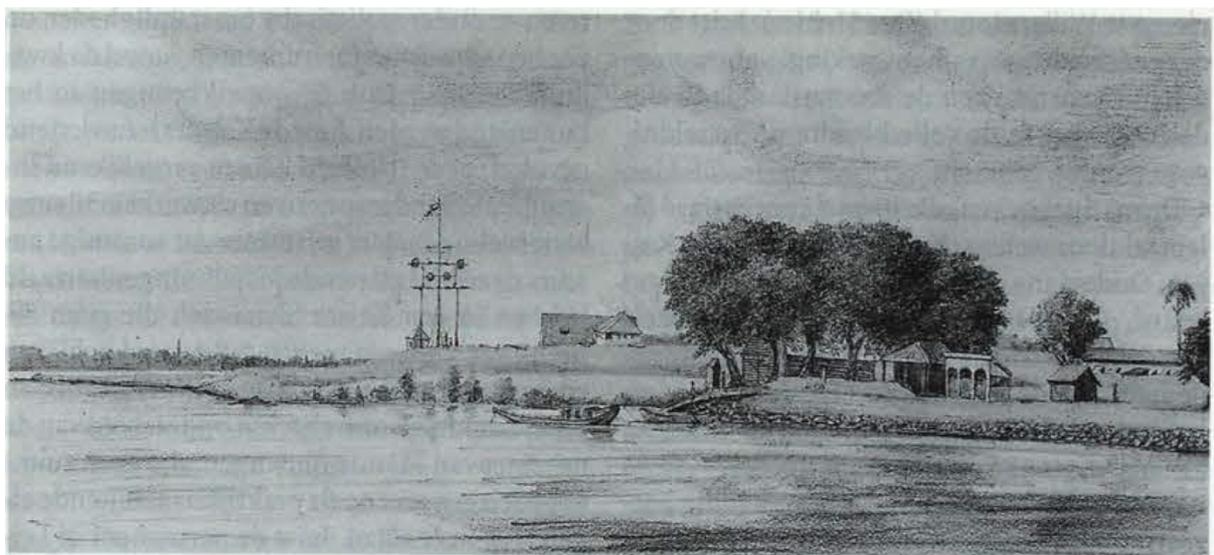

**Figure 3.** Time ball at Paramaribo. Drawing by H.A.H.M. Borret, 1878. (Collectie Koninklijk Instituut voor Taal-, Land- en Volkenkunde te Leiden, nr. 37A212; out of copyright.)

As a case in point, Cornelis van Schaick (1808–1874) wrote in his 1866 novel *De Manja: familie-tafereel uit het Surinaamse volksleven* ('The Manja: A Family Tableau from Surinamese Folk Life'; own translation): "The time ball on the guard ship was dropped, and at the very same moment, the noon gun fired, echoing thunderously through the woods on the far side of the river" (van Schaick, 1866: 10). Drawings such as Arnoldus H.A.H.M. Borret's (1848–1888) 1878 depiction of the city (see Figure 3) confirm the visual prominence of the time ball and fort. The apparatus became a landmark, a symbol of modernity on the Suriname River.

The apparent simplicity of this system masked disputes, however. In 1867, the commander of the guard ship *Kijkduin* (also rendered as *Kykduin*) proposed reducing the number of daily shots from three to two, arguing that regulations required only two firings in twenty-four hours (Oudschans Dentz, 1927/8). Governor Van Sypesteyn disagreed. He insisted that the full regimen of three shots—morning, noon and evening—be maintained, citing the population's attachment to the signals.

A compromise was reached. On 8 August 1867, Van Sypesteyn issued a resolution declaring that if naval vessels were unable to fire three times, Fort Zeelandia would fire the third shot (Oudschans Dentz, 1927/8: 441). In this dispute, timekeeping became a medium through which authority was negotiated. Naval officers viewed the signals as technical aids; the governor understood them as civic rituals and symbols of colonial order. Time itself thus became political.

**2.4 Later developments: Discs, flags and budgetary cuts**

By the late nineteenth century, Paramaribo's apparatus evolved in line with broader Dutch and international practice. Hydrographic manuals from the 1890s describe a black-and-white disc, 40.6 cm in diameter and 91.4 cm long, hoisted on the guard ship's main yard (U.S. Hydrographic Service, 1894: 35). Unlike the rotating flaps used in Amsterdam and elsewhere in the Netherlands, this disc was a local improvisation (Hite, 2021).

Flag signals were reported in 1890 and again between 1911 and 1916 (Kinns, 2022). By 1916 "… a cylinder with a red flag was dropped from a flagstaff on a building of the colonial government wharf at noon, local mean time …" (U.S. Hydrographic Service, 1916: 128). By 1912, however, official Dutch notices in 1912 warned mariners that the signal was "… unreliable and should not be used for the comparison of chronometers" (e.g., Hoofdkantoor van Scheepvaart, 1912; own translation). The U.S. Hydrographic Office echoed this in 1916: "The Netherlands Government has given notice that the time ball at Paramaribo, Dutch Guiana, is unreliable and should not be used for the comparison of chronometers." Budgetary pressures hastened this decline. In 1906 Governor Alexander W.F. Idenburg (1861–1935) abolished the morning and evening guns, retaining only the noon shot. His resolution read (Oudschans Dentz, 1927/8: 441–442; own translation):

> From 1 November 1906 the military complement at Fort Zeelandia will, instead of three, fire one time gun daily, at midday.
> 
> Paramaribo, 28 October 1906
> Captain Commander of the Garrison's Troops
> C[arel]. M[arinus]. H[ubert]. Kroesen [1861–1914]

This cut underscored the fiscal constraints of colonial administration. Where Batavia was sustained by telegraphy and observatories (e.g., Orchiston et al., 2021), Paramaribo's signals dwindled under budgetary retrenchment and limited maritime traffic. Despite their eventual decline, Paramaribo's time signals exemplify the hybrid character of colonial timekeeping. They were at once European and local. Their apparatus mirrored technological evolution in Europe—from balls to discs (flaps) to flags—yet their operation was shaped by tropical weather, fiscal constraint and colonial politics.

For mariners, the time signal assured accurate chronometers; for townspeople, it marked daily life. Oral traditions recall the noon gun as a cherished civic marker, remembered long after the apparatus itself disappeared. In this sense, Paramaribo's signals complicate conventional narratives. They were not mere copies of European prototypes, nor entirely autonomous local inventions. They were pragmatic compromises, blending naval practice, civic symbolism and imperial governance. Their story demonstrates how a global technology could acquire local meanings, becoming at once an instrument of navigation, a performance of authority and a shared ritual of urban life.

**3 DUTCH CARIBBEAN SIGNALS**

Paramaribo was not the only Atlantic outpost where Dutch authorities experimented with time signals. In the wider Caribbean, a handful of ports also attempted to provide visual or auditory signals for chronometer correction. The surviving documentation is fragmentary, scattered across hydrographic manuals, Dutch and British Admiralty notices and colonial records. Yet when pieced together, these fragments reveal a picture of pragmatic adaptation. Unlike Batavia or even Paramaribo, which attempted to replicate European routines in some detail, the Dutch Caribbean islands (Figure 4) operated simpler systems. Their signals reflected local conditions: the smaller scale of their harbours, the focus on inter-island trade and the limited willingness of colonial administrations to invest in costly apparatus.

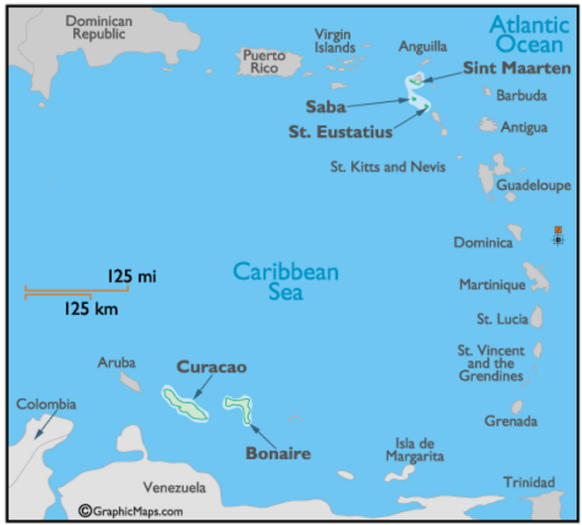

**Figure 4.** Dutch Caribbean islands geography.

The Dutch Antilles, comprising Curaçao, Aruba, Bonaire, St. Eustatius, Saba and St. Maarten, were strategically positioned along Caribbean trade routes. Curaçao functioned as the commercial and administrative centre of the Dutch West Indies. The island's natural deep-water ports, sheltered from open Atlantic swells, made it a strategic node for export of salt and dyewood from the colonial period (e.g., Debrot, 2009; Derix, 2016), and later for petroleum products. Smaller islands such as Aruba or Bonaire saw less extensive traffic, but nonetheless accommodated vessels engaged in intra-Caribbean and transatlantic voyages. As such, the need for navigational accuracy was present, although less acute than on the sandbank-strewn Surinamese coast.

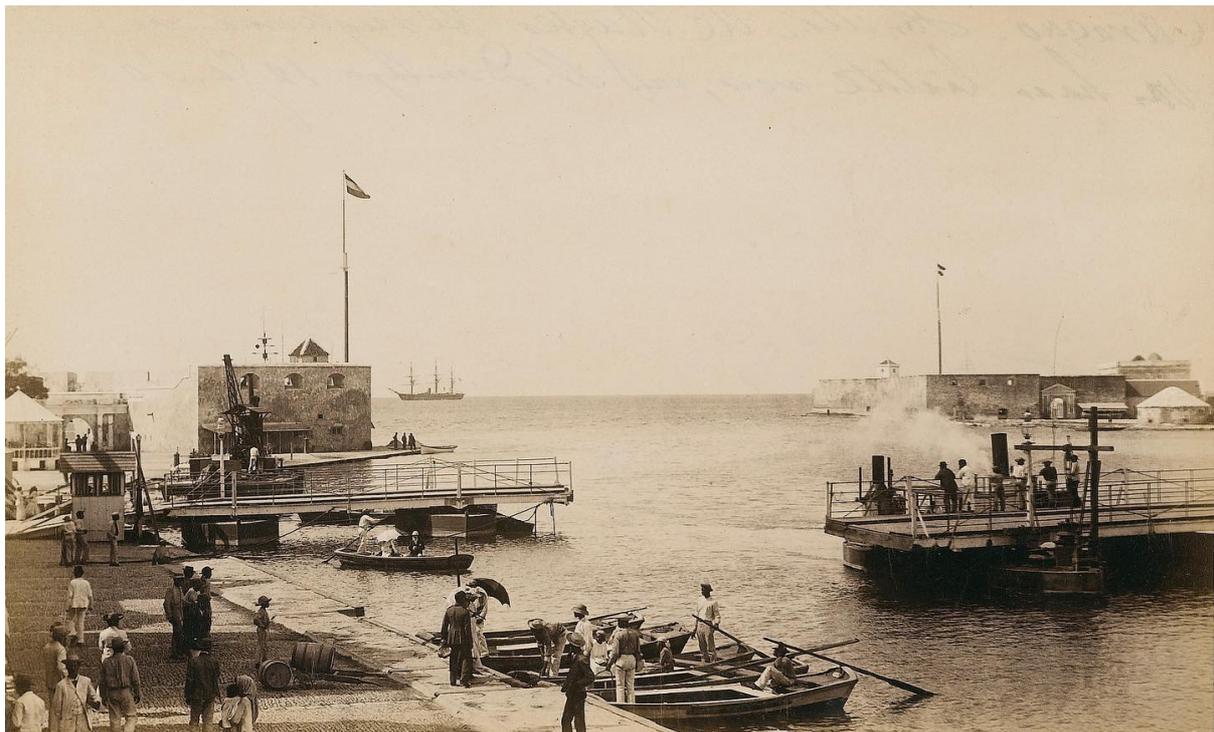

**Figure 5.** Harbour entrance to St. Anna Bay in Willemstad (Curaçao), showing the time flag on the left. (Photo: Soublette et Fils, 1893; out of copyright.)

### 3.1 Curaçao

The best-documented case is Curaçao. Hydrographic lists and Dutch/British Admiralty publications consistently record the island as maintaining a daily time signal from the late nineteenth century onwards. By the 1880s, Curaçao had adopted a time flag (see Figure 5), hoisted and lowered at a fixed hour each day (U.S. Hydrographic Service, 1894; see also Kinns, 2022). Curaçao's apparatus was deliberately modest. The choice of a flag was dictated by both practical and environmental considerations. It was inexpensive, easily maintained and well-suited to Curaçao's bright, dry climate.

In the harbour at Willemstad, where distances between anchorage and shore were short, a brightly coloured flag was clearly visible. Ships moored in St. Annabaai (St. Anna Bay) or Schottegat could readily observe the hoist and drop, without the need for larger or more complex apparatus.

This decision placed Curaçao within a broader regional pattern. Across the Caribbean, flag signals and noon guns were common in smaller ports. Hydrographic manuals compiled by the U.S. Hydrographic Office (1894, 1916) noted such practices in the British colonial outposts of St. Thomas (Virgin Islands), Port of Spain (Trinidad) and Bridgetown (Barbados), while a recent survey confirms that outside major centres like Kingston (Jamaica) and Havana (Cuba), these simple auditory or visual signals predominated (Kinns, 2022).

Contemporary practice suggests that simpler signals—flags or guns—were not uniquely Dutch or British but part of broader maritime practice in Western Europe. In metropolitan France, smaller ports sometimes deployed flag signals or other simple visual/time devices (for example at Brest and Saint-Nazaire) under French time-signal reforms from the 1850s onwards (Sauzereau, 2016a, 2016b). Evidence from Spain suggests similar pragmatic uses of guns and visual signals. The *Real Observatorio de la Armada* maintained time-ball practice in Cádiz and Madrid, and in the colonies cannon discharges were commonly used to mark hours. One well-documented case of an auditory time signal in the Spanish colonial world is Havana's *Cañonazo de las Nueve*, a nightly cannon shot at 9:00 p.m. that dates to the colonial era (Martínez de Baños Carrillo 2017; *La Habana: Fortalezas III*, 1931).

Curaçao's time flag thus represented an accommodation between Dutch standards and West Indian norms, a pragmatic solution suited to the scale and character of the harbour. Documentation from the early twentieth century confirms continuity of the practice. British Admiralty notices issued between 1911 and 1916 record Curaçao as maintaining a daily flag signal at Willemstad, noted in the *Notices to Mariners* and reprinted in the *Nautical Magazine* (British Admiralty, 1911–1916; *Nautical Magazine*, 1912, 1916). The daily time signal thus situated the island within a global constellation of maritime modernity, even if its apparatus was modest. To an arriving mariner, the daily hoist and drop signified that Curaçao, too, participated in the synchronised world of precision timekeeping.

The flag also carried symbolic weight. In Paramaribo, the noon gun had woven timekeeping into civic ritual; in Curaçao, the flag underscored the visual culture of maritime communication. Against the backdrop of colourful merchant flags in the bay, the daily time flag was a visible reminder of colonial order, reinforcing authority through routine.

### 3.2 Astronomical practice in the Caribbean

Whereas Paramaribo's naval officers engaged in sustained astronomical observation to determine local noon, Curaçao and the smaller Dutch Caribbean islands followed looser routines. The flag at Willemstad was not tied to daily sextant work by trained hydrographers but to approximate determinations of local time, often drawn from a single meridian altitude or even from pre-calculated tables (British Admiralty, 1911–1916). Curaçao's climate, dominated by bright skies and relatively predictable weather, reduced the observational difficulties faced in Suriname, but the smaller scale of shipping meant there was less incentive to pursue the highest astronomical precision.

The reliance on flags and occasional guns reflects this pragmatism. A noon gun fired from a fort could be based on a rough estimate of the Sun's meridian passage, sufficient for local or inter-island voyages where a few minutes of error in longitude posed little risk. For mariners on longer voyages, the practice was to rely on chronometer checks at larger ports—Paramaribo, Kingston or Havana—where more rigorous astronomical routines were maintained. Dutch and British Admiralty lists confirm this division of labour, registering Curaçao's flag as a recognised but secondary point of comparison.

Even so, the Caribbean signals were not devoid of astronomical content. They depended on the same celestial framework—the Sun's meridian passage, the equation of time, the offset from Greenwich—but translated it into a more economical routine. The daily hoist and drop of the flag thus embodied a simplified astronomical adaptation of European precision to colonial circumstances, adequate for local trade but always connected, at least symbolically, to the global astronomical order.

### 3.3 Other islands

Evidence for permanent time signals on the smaller Dutch Caribbean islands is far thinner. Hydrographic manuals and British or Dutch Admiralty notices do not record fixed time-signal apparatus at Aruba, Bonaire or St. Eustatius. Their silence is revealing: whereas Curaçao's daily flag signal was noted in both the *Notices to Mariners* and the *West Indies Pilot*,[5] no such entries exist for the other Dutch islands. In practice, mariners there relied on occasional signals from naval guard ships or on the long-established custom of noon guns fired from forts (Howse, 1980; Kinns, 2024). At St. Eustatius, for example, fortifications dating to the eighteenth century occasionally fired salutes or noon guns. Yet there is no evidence of a systematic time ball or flag. The same holds for Aruba and Bonaire, where shipping traffic was largely inter-island. For vessels sailing only short distances, the need for precise chronometer correction was less pressing than for transoceanic departures. Auditory signals sufficed for rough checks, even if they lacked the precision of balls or discs.

### 3.4 Comparison with Paramaribo

Placing Curaçao and the smaller Dutch Caribbean islands alongside Paramaribo highlights the variety of Dutch colonial temporal regimes. In Suriname, the capital's role as a transatlantic hub justified sustained investment in a hybrid system of ball, gun, disc and flag. Mariners departing for Europe depended on accurate chronometer checks to avoid catastrophic errors when approaching shallow coasts. Civic authorities, too, embraced the noon gun as a marker of urban routine.

Curaçao, by contrast, maintained a simpler flag signal, sufficient for regional inter-island trade and occasional transatlantic traffic. Its apparatus was cheap, reliable and suited to local conditions. The island's prominence as a trading hub made some form of signal desirable, but the demands of precision were lower than in Suriname. On Aruba, Bonaire and St. Eustatius, the absence of fixed apparatus reflects their marginal role in global shipping. For local sailors, rough noon checks sufficed; for visiting vessels, chronometer comparisons could be made at larger ports such as Curaçao/Willemstad. The reliance on occasional guns underscores the flexible, *ad hoc* character of Dutch practice.

These differences also highlight the decentralised nature of Dutch colonial administration. Unlike Britain, which sought to standardise time signals across its empire, the Netherlands tolerated a patchwork of solutions. This was in part a matter of scale: the Dutch empire, while globally dispersed, was smaller and less financially powerful. It was also a matter of philosophy. Dutch hydrography was highly professionalised, but colonial governors exercised wide discretion in matters of expenditure. The result was a spectrum of temporal infrastructures: from Paramaribo's hybrid system, to Curaçao's modest flag, to the guns of the smaller islands.

Situating the Dutch Caribbean within wider imperial practice underscores both similarities and differences. The British installed time balls in Kingston and Havana as part of a global network tied to Greenwich. These installations testified to imperial ambition and the desire to display European authority in colonial capitals. The Dutch, by contrast, pursued a more modest approach. Batavia and Paramaribo were equipped with time balls, but no equivalent apparatus appeared in the Caribbean. Curaçao's flag was adequate but never spectacular. This difference reflected not only resources but also the Netherlands' secondary position in the global maritime order. By the nineteenth century, Britain set the standards; Dutch hydrography, while competent, adapted to them.

Nonetheless, the presence of any signals at all reveals Dutch participation in the global system. Hydrographic manuals published in London or Washington regularly listed Curaçao's flag and Paramaribo's disc or gun. Mariners navigating the Caribbean could thus treat Dutch ports as part of the same international infrastructure, even if their apparatus was less elaborate. The inclusion of these signals in British Admiralty and U.S. Hydrographic Office publications indicates that they were considered reliable enough, at least in principle, to be of use.

### 3.4 Civic reception

Documentation of civic reception in the Caribbean is sparser than in Suriname. There is little evidence that Curaçao's flag acquired the same communal resonance as Paramaribo's noon gun. Curaçao's signal was visual rather than auditory, limiting its impact on townspeople who were not directly observing the harbour. Its utility was primarily maritime, reinforcing the island's role as a trading hub rather than structuring civic temporality. Yet even here, symbolism mattered. In a harbour filled with the colours of merchant vessels from across the Atlantic world, the official time flag was a reminder that the Netherlands retained sovereign control. Its routine performance linked Curaçao, however modestly, to the broader theatre of maritime modernity.

Taken together, the Dutch Caribbean islands illustrate a patchwork of temporal regimes. Paramaribo pursued a hybrid system, anchored in both naval science and civic ritual. Curaçao adopted a modest flag, effective for its scale and suited to its climate. The smaller islands relied on guns, adequate for local needs. This patchwork reflects the pragmatism of Dutch colonial practice. Rather than impose uniform standards, the Netherlands tolerated diversity. Each island adopted the minimal apparatus necessary for its shipping traffic and fiscal resources. This flexibility ensured that Dutch Caribbean ports were not excluded from global navigational networks, but it also meant that they never projected the same authority as Greenwich, Cape Town or Sydney.

# 4 DISCUSSION

In the preceding sections I have traced the establishment and evolution of Dutch colonial time signals in Paramaribo and the Caribbean. What emerges is a story not of uniform diffusion from the metropole but of pragmatic adaptation, contested authority and hybrid infrastructures. In this section, I will draw together the key themes of material adaptation, political negotiation, maritime function and imperial context. These themes illuminate how the Dutch Atlantic colonies participated in global regimes of precision time while also reshaping them according to local conditions.

## 4.1 Material adaptation and technical improvisation

One of the most striking features of the Dutch colonial apparatus is its diversity. Unlike the Netherlands itself, where Leiden Observatory supplied astronomical authority for balls from Den Helder in the north to Vlissingen in the south (see my companion article), the colonies lacked local scientific institutions of comparable standing. Hydrographic officers aboard naval guard ships thus became the *de facto* astronomers of the tropics, determining local noon with sextants, artificial horizons and patient observations under difficult conditions. In Paramaribo, this produced a layered sequence of apparatus: a naval time ball, a noon gun at Fort Zeelandia, a disc on the guard ship and a flag or cylinder from the government wharf. Curaçao relied on a flag; smaller islands on occasional cannon fire. The evolution of these devices shows both continuity with European practice and inventive improvisation. The discs described in late nineteenth-century manuals, for example, were peculiar to Paramaribo, not replicas of the rotating shutters or flaps used in Amsterdam (Hite, 2021).

This material diversity reflected environmental factors. In humid Suriname, wooden mechanisms deteriorated quickly, demanding replacements and improvisation. In sunny, compact Curaçao, a flag sufficed. Such adaptation was typical of colonial technology. As historians of science have emphasised, instruments transplanted from Europe were rarely used unchanged; they were modified to suit new climates, materials and needs (Schaffer et al., 2009). Dutch colonial time signals were no exception. Time signals were not merely technical devices; they were also political instruments. In Paramaribo, Governor Van Sypesteyn's intervention in 1867 illustrates how temporal authority could become a matter of governance. When naval officers aboard the *Kijkduin* proposed reducing the number of daily gun shots, the governor insisted that the population valued the full regimen of three. The compromise—Fort Zeelandia firing the third shot if necessary—embedded the signal more firmly in civic routine.

Here we see timekeeping as a focus of power. Naval officers viewed the signals in terms of hydrographic utility; the governor saw them as civic ritual. The population clearly oriented daily life around the noon gun. The signal thus operated simultaneously on technical, political and social registers. In Curaçao, the political dimension was less contested but no less present. The daily hoist and drop of a Dutch flag (cf. Figure 5) underscored colonial sovereignty in a harbour dominated by foreign shipping. The flag signal was an assertion of Dutch presence, a reminder that even in a remote Caribbean port the rhythms of maritime life were still anchored in the authority of the European colonial masters.

## 4.2 Maritime function and global navigation

The primary justification for colonial time signals remained their maritime utility. Chronometers were fragile instruments. Subject to temperature fluctuations and mechanical wear, they required regular comparison against reliable local time. A visible or audible noon signal allowed captains to reset or at least verify their instruments before embarking on long voyages.

In Paramaribo, this function was critical. The Suriname River estuary was notoriously hazardous, with shifting sandbanks and treacherous bars. By checking their chronometers against the noon ball or gun, mariners departing for Europe reduced the risk of stranding. Hydrographic surveys by Dutch officers such as Jan Carel Pilaar (1798–1849) and Swart reinforced this function, producing charts whose accuracy depended on reliable temporal frameworks.

In Curaçao, the function was more modest but still significant. Ships engaged in intra-Caribbean trade or transatlantic crossings could use the daily flag to confirm their instruments' accuracy. British and Dutch Admiralty lists consistently recorded Curaçao's signal, indicating that it was considered trustworthy enough to be of practical use. Even the smaller islands, with their occasional guns, provided mariners with at least rough checks. The maritime function of these signals linked Dutch colonies to a global system. Hydrographic manuals published in London and Washington routinely listed Paramaribo and Curaçao alongside Cape Town, Sydney and San Francisco. Mariners consulting such guides would see Dutch ports inscribed within the same temporal geography as those of Britain, France or the United States. Even modest signals thus had global resonance.

The Dutch colonial case also reveals how empires mediated science through hierarchies of importance. Batavia, capital of the East Indies, received greater investment: a land-based time ball linked to an observatory, later supplemented by telegraphic signals. Paramaribo, although significant, made do with naval observations and somewhat improvised apparatus. Curaçao relied on a flag. Smaller islands received nothing permanent. This hierarchy reflected both economic priorities and imperial ideology. The East Indies, the jewel of the Dutch empire, merited more elaborate infrastructures of science. The Atlantic colonies, by contrast, were peripheral, their plantations less lucrative, their trade less central. The modesty of their time signals reflected their place in the imperial hierarchy.

Yet even in this hierarchy, the Atlantic colonies mattered. They demonstrated Dutch participation in the global maritime system; they provided indispensable waypoints for ships crossing the Atlantic. Their signals, however small, anchored Dutch sovereignty in a crowded imperial arena. The Netherlands may have ceded primacy to Britain, but it continued to perform scientific authority in its own colonial sphere.

### 4.3 Broader implications: Temporality, modernity and decline

The Dutch Atlantic signals also speak to broader histories of temporality and modernity. They exemplify 'time discipline'—the structuring of daily life through mechanical precision (Thompson, 1967). In Paramaribo, the noon gun regulated civic routine; in Willemstad, the daily flag asserted maritime order. These signals tied the Dutch colonial towns into global rhythms, synchronising them with European standards. At the same time, their decline illustrates the fragility of colonial infrastructures. By the early twentieth century, Paramaribo's signals were declared unreliable; Curaçao's flag continued but with diminishing importance. Wireless telegraphy and radio time signals, introduced from the 1910s (e.g., Howse, 1980), rendered visual and auditory signals increasingly obsolete. Ships could now receive time directly via radio, removing the need for local balls or guns. Budgetary pressures accelerated this transition. In 1906, Governor Idenburg abolished the morning and evening guns in Paramaribo, retaining only the noon shot. In Curaçao, the flag continued into the 1910s but never developed beyond its modest form. By the interwar years, Dutch Atlantic time signals had effectively disappeared, replaced by wireless alternatives.

Bringing these themes together, we can see Dutch colonial time signals as hybrid infrastructures, combining European models, local improvisations, political authority and maritime function. They were never simple transplants of European technology. Instead, they were adapted to tropical climates, constrained by budgets, contested by governors and naval officers, and embedded in civic life. Their history challenges linear narratives of scientific diffusion. Rather than a unidirectional flow from Europe to the colonies, we see a process of negotiation, adaptation and improvisation. Colonial time signals were not merely tools of navigation; they were also performances of authority, rituals of community and symbols of empire.

## 5 CONCLUSION

The history of Dutch colonial time signals in the Atlantic world reveals the uneven geography of modernity. In Paramaribo and Curaçao, as in other colonial ports, the apparatus of precision timekeeping took root in distinctive forms: balls and guns, discs and flags, cylinders and cannons. These were modest devices, improvised from available materials and operated under local conditions. Yet they anchored Dutch colonial towns in a global system of navigation and temporal discipline, linking them—however tenuously—to the great centres of maritime modernity.

Three themes stand out from this survey:

1. **Material diversity.** Unlike the standardised installations of Greenwich or Washington, Dutch colonial signals were eclectic. In Paramaribo, the apparatus shifted from naval time balls to shore-based guns, then to discs, flags and cylinders. In Curaçao, a simple flag sufficed. On smaller islands, only occasional guns marked the hour. This diversity reflects the environmental challenges, fiscal limitations and improvisational ingenuity of colonial administration. The colonial signal was rarely the metropolitan signal transplanted intact; it was a hybrid, adapted to local needs and constraints.

2. **Political negotiation.** Time signals were not only technical but also political. In Paramaribo, Governor van Sypesteyn's dispute with naval officers over the number of daily shots reveals how civic expectations could shape maritime practice. In Curaçao, the daily hoist and drop of a Dutch flag signalled sovereignty in a harbour filled with foreign shipping. This underscores how technologies of time became entangled with authority, legitimacy and identity. To control the hour was to control the rhythms of colonial life.

3. **Maritime function and imperial hierarchy.** At their core, these signals served mariners, providing points of comparison for fragile chronometers. They reduced the risks associated with transatlantic navigation and embedded Dutch colonies into international hydrographic manuals. Yet their form and reliability also reflected imperial hierarchies. Batavia, jewel of the East Indies, received an observatory and telegraphic signals. Paramaribo made do with naval observations and improvised apparatus. Curaçao adopted a modest flag; smaller islands, only guns. The distribution of temporal infrastructure mirrored the economic and political priorities of the empire.

Together, these themes complicate the notion of colonial science as simple diffusion. The Netherlands did not simply export European apparatus to the Atlantic. Instead, it created patchwork infrastructures, negotiated between naval officers, governors and local populations, adapted to climate and budget, and inscribed in civic as well as maritime life. These infrastructures were simultaneously scientific and social, technical and political. They were instruments of navigation, but also of sovereignty and community.

The Dutch Atlantic case also offers comparative insight. It demonstrates that maritime modernity was not uniform, nor exclusively British. While Greenwich set the standard, smaller empires improvised their own ways of keeping time at sea. Dutch colonial signals were modest, yet they inserted Paramaribo, Curaçao and other ports into the same global temporal framework as Kingston, Cape Town or Sydney. They reveal how even minor colonial outposts participated in, and contributed to, the synchronisation of global navigation.

For historians of science and empire, this case suggests several broader implications. First, it highlights the importance of studying seemingly ordinary technologies. The time ball or flag may lack the grandeur of observatories or telegraph lines, but they played vital roles in the circulation of knowledge and authority. Second, it underscores the hybridity of colonial infrastructures: neither wholly European nor wholly local, but pragmatic compromises shaped by negotiation and adaptation. Third, it shows how technologies of time were inseparable from politics. To drop a ball or fire a gun was to enact authority, regulate routine and inscribe empire in the daily lives of colonial residents.

Finally, the Dutch Atlantic reminds us of the unevenness of modernity itself. Precision timekeeping spread across the globe not as a uniform wave but as a patchwork quilt. Some places received observatories; others, flags. Some heard the noon gun daily; others relied on occasional shots. The result was a differentiated temporal landscape, in which Paramaribo and Curaçao were modest but significant nodes.

In bringing this history to light, we gain a more nuanced understanding of both Dutch colonial science and global maritime modernity. Time signals in the Atlantic were neither peripheral curiosities nor footnotes to Greenwich. They were central to the lived experience of sailors and townspeople alike, vital to the functioning of navigation and emblematic of the ways in which empires inscribed themselves in the fabric of daily life. They remind us that modernity was not only measured in observatories and telegraphs but also in simple devices on harbour masts and fortress walls. Through them, Dutch colonial authorities projected order, discipline and belonging—however fragile, however improvised—into the uncertain world of the nineteenth-century Atlantic.

# 6 NOTES

1. The U.S. Hydrographic Service manuals of both 1894 and 1916 imply that the ball (or disc) was hoisted to the top three minutes before the reference time. The 1894 manual also includes a slightly different time difference for the guard ship, 3h 40m 39.7s (by 1916 this had changed to 3h 40m 35.2s).
2. Although J. Vos is referenced in contemporary publications, I have been unable to identify this officer given that 'Vos' is a common Dutch surname.
3. This was the Amsterdam-based Danish clockmaker Andreas Hohwü (1803–1885).
4. Dutch mariners often relied on the British *Nautical Almanac*. However, several equivalent publications in Dutch appeared at various times in the eighteenth and nineteenth centuries, including the *Zee- en Sterrekundige Almanak* ('Nautical and Astronomical Almanac'), the *Nederlandsche Sterrekundige Almanak* ('Dutch Astronomical Almanac'), *Sterrekundige Jaarboeken* ('Astronomical Yearbooks') and Dutch Hydrographic Office tables.
5. The British Admiralty's *West Indies Pilot* (7th ed., 1920) describes Curaçao's harbour facilities and records a routine daily time flag at Willemstad, while the island entries for Aruba, Bonaire and St.

Eustatius give anchorage, depth and landmark information but make no mention of fixed time-signal apparatus (British Hydrographic Department, *West Indies Pilot*, Vol. II, 7th ed., 1920; U.S. Hydrographic Office, *Sailing Directions for the West Indies*, late-19th–early-20th centuries). The Admiralty's weekly *Notices to Mariners* (1911–1916) supply the contemporary notices on Curaçao's flag time signal reproduced in the *Pilots*. See also Kinns (2024) for a modern survey of Admiralty signal lists.

# 7 REFERENCES

Bartky, I.R., and Dick, S.J., 1981. The First Time-Balls. *Journal for the History of Astronomy*, 12, 155–164.
British Admiralty, 1911–1916. *Notices to Mariners*. London, Hydrographic Department. Each year's bound volume includes Curaçao's daily time signal under 'Foreign Time Signals'.
British Hydrographic Department (Admiralty), 1920. *The West Indies Pilot,* vol. II (7th ed., J. G. Boulton, ed.). London, Hydrographic Department. See entry for Curaçao/Willemstad' compare with entries for Aruba, Bonaire and St. Eustatius, which contain harbor particulars but no time-signal listing.
Debrot, A.O., 2009. *Cultural ties to the land in an arid plantation setting in Curaçao*. Final report submitted to Commission of the Island Territory of Curaçao. Nomination to UNESCO of a Cultural Landscape. Curaçao, CARMABI Foundation; https://www.researchgate.net/publication/311811250_Cultural_ties_to_the_land_in_an_arid_plantation_setting_in_Curacao
de Grijs, R., 2017. *Time and Time Again: Determination of Longitude at Sea in the 17th Century*. Bristol UK, Institute of Physics Publishing.
Derix, R.R.W.M., 2016. *The history of resource exploitation in Aruba*. Part of the series 'Spatial Developments in the Aruban Landscape: A multidisciplinary GIS-based approach derived from geologic, historic, economic and housing information', vol. 2. Aruba, Central Bureau of Statistics. https://cbs.aw/wp/wp-content/uploads/2016/12/H2_LandscapeSeries_The_History_Of_Resource_Exploitation_In_Aruba_10-10-2016-1.pdf.
Gillin, E.J., 2020. Tremoring transits: railways, the Royal Observatory and the capitalist challenge to Victorian astronomical science. *British Journal for the History of Science*, 53, 1–24.
Glennie, P., and Thrift, N., 1996. Reworking E. P. Thompson's 'Time, Work-Discipline and Industrial Capitalism'. *Time & Society*, 5(3), 275–299.
Hite, L.L., 2014, updated 2021. *How Time Balls Worked, featuring the Cincinnati Observatory*; https://prancer.physics.louisville.edu/modules/time/articles/how_time_balls_worked.pdf.
Hoofdkantoor van Scheepvaart, 1912. *Bericht aan Zeevarenden*, No. 17(130). 's Gravenhage. Cited by Kinns (2022).
Howse, D., 1980. *Greenwich Time and the Longitude*. London, Philip Wilson.
Kinns, R., 2021. Time Signals for Mariners in SE Asia: Time Balls, Discs, Bells, Guns and Lights (Chapter 12). In: Orchiston, W., and Vahia, M.N. (eds), *Exploring the History of Southeast Asian Astronomy*, Cham, Springer: 411–460.
Kinns, R., 2022. Visual time signals for mariners between their introduction and 1947: A new perspective. *Journal of Astronomical History and Heritage*, 25(4), 601–713.
Kinns, R., 2024. *The World of Visual Time Signals for Mariners: Time Balls, Time Guns, Time Lights and Other Signals.* Cham, Springer.
*La Habana: Fortalezas III. El cañonazo de las 9: Un sevillano que da la hora.* (18 December 1931) (in Spanish). *Biblioteca Histórica Cubana y Americana 'Francisco González del Valle'* (Emilio Roig de Leuchsenring collection); https://repositoriodigital.ohc.cu/s/repositoriodigital/item/6586.
Landes, D.S., 1983. *Revolution in Time: Clocks and the Making of the Modern World*. Cambridge MA, Harvard University Press.
Martínez de Baños Carrillo, F., 2017. El cañonazo de las 9 (in Spanish). *Armas y Cuerpos*, 136, 91–97; https://publicaciones.defensa.gob.es/media/downloadable/files/links/r/e/revista_ac_136.pdf.
Milne, J., 1899. Civil Time. *The Geographical Journal*, 13(2), 173–194.
Mohrmann, J.M., 2003. De Koninklijke Marine als vernieuwer van de zeevaartkunde, 1850–1900 (in Dutch). *Tijdschrift voor Zeegeschiedenis*, 22(1), 44–58.
Notice to Mariners, No. 130 (1912): Curaçao—Time Signal. *Nautical Magazine*, 6.
Notice to Mariners, No. 217 (1916): Curaçao—Time Signal. *Nautical Magazine*, 221.
Orchiston, W., Sungging Mumpuni, E., and Steinicke, W., 2021. J.A.C. Oudemans and Nineteenth Century Astronomy in the Dutch East Indies. In: Orchiston, W., and Vahia, M.N. (eds), *Exploring the History of Southeast Asian Astronomy*, Cham, Springer: 285–316.

Oudschans Dentz, F., 1927/8. Hoe men in Suriname in den loop der eeuwen den tijd heeft aangegeven (in Dutch). *De West-Indische Gids*, 9, 433–448.
Raj, K., 2007. *Relocating Modern Science: Circulation and the Construction of Knowledge in South Asia and Europe, 1650–1900*. Basingstoke, Palgrave Macmillan.
Sauzereau, O., 2016a. Les signaux horaires français: la quête d'un système uniformisé (in French). *Entre Ciel et Mer*. Cahiers de l'Observatoire du Patrimoine et de la Mer.
Sauzereau, O., 2016b. French time signals: the quest for a standardized system. *Cahiers François Viète*, II-8/9, 179–202.
Schaffer, S., Roberts, L.L., Raj, K., & Delbourgo, J. (eds.), 2009. *The Brokered World: Go-Betweens and Global Intelligence, 1770–1820*. Sagamore Beach MA, Science History Publications.
Swart, J., 1853. *Verhandelingen en Berigten betrekkelijk het Zeewezen en de Zeevaartkunde* (in Dutch). Amsterdam, Wed. G. Hulst van Keulen.
Thompson, E.P., 1967. Time, Work-Discipline, and Industrial Capitalism. *Past & Present*, 38, 56–97.
U.S. Hydrographic Service, 1894. *East Coast of South America: From the Orinoco River to Cape Virgins*, 2nd ed. Washington DC, Government Printing Office.
U.S. Hydrographic Service, 1916. *South America Pilot, Vol. 1 (East Coast): From the Orinoco River to the Plata River*, 1st ed. (H.O. No. 172). Washington DC, Government Printing Office.
van Cleef, Gebr., 1858. III. Opgaven der breedte en lengte van eenige belangrijke plaatsen (in Dutch). In: *Almanak voor de Nederlandsche West-Indische bezittingen, en de kust van Guinea. Jaargang 1859*. Den Haag, De Gebroeders Van Cleef.
van Schaick, C., 1866. *De Manja: familie-tafereel uit het Surinaamse volksleven* (in Dutch). Arnhem, D. A. Thieme; https://www.dbnl.org/tekst/scha031manj01_01/index.php.
Wesenhagen, J.E., 1851. Berigt aan Zeevarenden (in Dutch). *Surinaamsche Courant en Gouvernements advertentieblad*, 13 November 1851. P. 1.